\documentclass[11pt]{iopart}

\usepackage{graphicx}
\usepackage{color}
\usepackage{epsfig}
\usepackage{float}
\usepackage{multirow}
\expandafter\let\csname equation*\endcsname\relax
\expandafter\let\csname endequation*\endcsname\relax
\usepackage{amsfonts}
\usepackage{amssymb}
\usepackage{amsmath}
\usepackage{tabularx,url,color}
\usepackage{hyperref}

\newcommand{\Neff}{\ensuremath{\mathrm{N_{eff}}}}
\newcommand{\lambdaCDM}{\ensuremath{\mathrm{\Lambda_{CDM}}}}
\newcommand{\hzero}{\ensuremath{\mathrm{h_{0}}}}
\newcommand{\Omegagwhdeux}{\ensuremath{\mathrm{\Omega_{GW}h_{0}^{2}}}}
\newcommand{\Omegagw}{\ensuremath{\mathrm{\Omega_{GW}}}}

\begin{document}
\leftline{Dated: \today}

\title{Improved constraint on the primordial gravitational-wave density using recent cosmological data and its impact on cosmic string models}

\begin{abstract}
The production of a primordial stochastic gravitational-wave background by processes
occuring in the early Universe is expected in a broad range of models. Observing
  this background would open a unique window onto the Universe's
  evolutionary history. Probes like the Cosmic Microwave Background (CMB) or
  the Baryon Acoustic Oscillations (BAO) can be used to set upper
  limits on the stochastic gravitational-wave background energy density $\Omegagw$
  for frequencies above $10^{-15}$~Hz. We perform a profile likelihood analysis of the
  Planck CMB temperature anisotropies and gravitational lensing data combined with 
  WMAP low-$\ell$ polarization, BAO, South Pole Telescope and Atacama Cosmology 
  Telescope data. We find that $\Omegagwhdeux < 3.8 \times 10^{-6}$ at
  95\% confidence level for adiabatic initial conditions which improves over the 
  previous limit by a factor 2.3. Assuming that the primordial gravitational waves 
  have been produced by a network of cosmic strings, we have derived exclusion 
  limits in the cosmic string parameter space. If the size of the loops is determined 
  by gravitational back-reaction, string tension values greater than 
  $\sim 4 \times 10^{-9}$ are excluded for a reconnection probability of $10^{-3}$.
\end{abstract}

\author{
  Sophie Henrot-Versill\'e$^1$, 
  Florent Robinet$^1$,  
  Nicolas Leroy$^1$, 
  St\'ephane Plaszczynski$^1$, 
  Nicolas Arnaud$^1$, 
  Marie-Anne Bizouard$^1$,
  Fabien Cavalier$^1$, 
  Nelson Christensen$^{1,2}$, 
  Fran\c{c}ois Couchot$^1$, 
  Samuel Franco$^1$, 
  Patrice Hello$^1$, 
  Dominique Huet$^1$, 
  Marie Kasprzack$^1$, 
  Olivier Perdereau$^1$,
  %Benjamin Rouill\'e d'Orfeuil$^1$, 
  Marta Spinelli$^1$,
  Matthieu Tristram$^1$
}

\address{$^1$LAL, Univ Paris-Sud, CNRS/IN2P3, Orsay, France}
\address{$^{2}$Physics and Astronomy, Carleton College, Northfield, MN  55057, USA}

%\pacs{11.27.+d, 98.80.Cq, 11.25.-w}
\maketitle

\section{Introduction} \label{sec:introduction}

A stochastic background is expected to form from the incoherent 
superposition of a large number of gravitational wave (GW) signals
emitted by many sources of astrophysical and cosmological origin.
Examples of GW astrophysical sources are compact binary coalescence, 
core collapse supernovae or rotating neutron stars~\cite{Wu:2011ac,Wu:2013xfa,Marassi:2009ib}. 
Among the cosmological 
sources, many GW generation mechanisms have been proposed covering a broad 
range of frequencies, such as the amplification of quantum vacuum 
fluctuations during inflation~\cite{Grishchuk:1974ny,Starobinski:1979aa,BarKana:1994bu,Easther:2006vd,Lopez:2013mqa}, 
first order phase transitions~\cite{Apreda:2001us,Leitao:2012tx}, 
cosmic strings~\cite{Kibble:1976sj,VilenkinShellard:94,Olmez:2010bi} 
and pre Big-Bang models~\cite{Brustein:1995ah,Buonanno:1996xc}. 
These GWs provide a unique probe of the evolution of the Universe 
from its birth as they travel through space-time with virtually no interaction 
with matter.

%Many mechanisms predict the existence of a stochastic cosmological
%gravitational-wave (CGW) background \cite{Maggiore:1999vm} over a
%large range of frequencies, such as, for instance, quantum
%fluctuations during the
%inflation~\cite{BarKana:1994bu,Easther:2006vd,Lopez:2013mqa},
%electroweak phase transitions~\cite{Leitao:2012tx}, or cosmic
%strings~\cite{Kibble:1976sj,VilenkinShellard:94,Olmez:2010bi}. 

The stochastic background of GWs is described
in terms of its energy spectrum as function of the frequency:
\begin{equation}
\Omegagw(f)={d\Omegagw}/{d(\ln{f})} \ ,
\end{equation}
where $\Omegagw$ is the total
energy density of GWs relative to the critical energy density. This
spectrum is constrained for specific frequency ranges by direct
GW searches. Assuming a flat spectrum, the LIGO and Virgo collaborations 
recently set the limit $\Omegagw(f\simeq 100\ \text{Hz})\times (\hzero/0.68)^2<5.6\times
10^{-6}$~\cite{Aasi:2014zwg}, where $\hzero=H_0/(100$~km~s$^{-1}$~Mpc$^{-1})$ 
is the reduced Hubble parameter. 
Pulsar timing experiments provide a constraint at much lower frequencies: 
$\Omegagw(f=2.8\ \text{nHz})\times (\hzero/0.73)^2
< 1.3 \times 10^{-9}$~\cite{Shannon:2013wma}. These  constraints
apply to any GW background, both of cosmological and astrophysical origin. 

As far as cosmological gravitational waves (CGW) are concerned, they could ultimately be detected,
if generated by inflation
models, through the measurement of the tensor-to-scalar ratio
with low-$\ell$ Cosmic Microwave Background (CMB) B polarization anisotropies~\cite{Kamionkowski:1997av,PhysRevD.55.R435}.  Still indirect
bounds on their energy density can be set from different cosmological probes. 
For instance, Big Bang Nucleosynthesis (BBN), 
through the measurement of the light element abundances gives
$\Omegagwhdeux < 8.1 \times 10^{-6}$ for all frequencies above $10^{-10}$ Hz~\cite{Allen:1996vm,Maggiore:1999vm,Cyburt:2004yc}. 
CMB and  Baryon Acoustic Oscillation 
(BAO) data can also be used to set limits on a CGW background energy density for
frequencies greater than $10^{-15}$ Hz~\cite{Smith:2006nka,Sendra:2012wh}.

In this article, we have revisited the constraints on $\Omegagwhdeux$ using the 
high-precision measurements of the CMB properties from the Planck 
collaboration~\cite{Ade:2013kta,Ade:2013tyw} combined with up-to-date BAO data~\cite{Anderson:2013oza}.
The study is made within the 
$\lambdaCDM$ scenario which has proved to be able to successfully describe a wide
range of cosmological data~\cite{Ade:2013kta}. In order to be as model independent as possible, we have 
considered the case for which the CGWs are produced under adiabatic conditions.
Within this framework, the influence of CGWs on the CMB power spectrum is 
expected to be the same as extra massless neutrino species~\cite{Smith:2006nka,Sendra:2012wh}. 
We then perform a profile likelihood analysis on the effective number of relativistic degrees of freedom, 
$\Neff$, with different hypotheses on the sum of neutrinos masses 
($\sum{m_\nu}$), and derive the corresponding upper limits on $\Omegagwhdeux$. The limit 
derived with $\sum{m_\nu}$ free in the fit is thereafter used to constrain cosmic string model parameters assuming the CGW background 
was produced by a network of cosmic strings.

In section~\ref{sec:sgwbg}, we briefly present the theoretical framework used
to constrain the CGW background energy density. The 
description of the cosmological data, the statistical procedure
and the results on $\Omegagwhdeux$
are discussed in Section~\ref{sec:planck}. The impact on cosmic 
string models is described in Section~\ref{sec:cs}. Finally we discuss in
Section~\ref{sec:conclusions} the future prospects. 
Throughout this paper, we take the speed
of light and the reduced Planck constant equal to 1.

%The measurement of the
%effective number of relativistic species, $\Neff$, is then
%translated into constraints on the total CGW energy density,
%integrated over all frequencies:
%\begin{equation} \label{eq:omega}
%  \Omegagw = \int_{f_{min}}^{\infty}{\Omegagw(f)}{d(\ln{f})} \ .
%\end{equation}

%Until this publication, the tightest bound was obtained from the
%analysis of CMB data up until the WMAP 7-year data
%release~\cite{Sendra:2012wh}. The analysis presented in this paper
%includes the high-precision measurements of the
%CMB spectrum from the Planck satellite~\cite{Ade:2013kta,Ade:2013tyw} combined
%with up-to-date BAO data. 

%In this article we consider the case where the extra relativistic
%degrees of freedom are at least contained in the neutrino and the
%CGW components. We only
%consider the case for which the CGW background energy-density
%perturbations are adiabatic: this is motivated by the fact
%that adiabatic cosmic gravitational waves with wavelengths shorter
%than the sound horizon at decoupling behave as free streaming massless
%particles~\cite{Sendra:2012wh}. The limit on the CGW density from
%CMB and BAO data derived below extends to low frequencies of the
%order of $10^{-15}$ Hz~\cite{Maggiore:1999vm,Smith:2006nka}.

\section{The stochastic gravitational-wave background density} \label{sec:sgwbg}

%The early Universe can be described by a radiation-dominated equation
%of state. 
The radiation energy density relative to the critical density,
$\Omega_{rad}$, can be written as the sum of the relativistic
contributions of photons ($\gamma$), neutrinos ($\nu$), and
any possible extra radiation ($x$):
\begin{equation} \label{eq:rhorad}
  \Omega_{rad}=\Omega_{\gamma}+\Omega_{\nu}+\Omega_x \ .
\end{equation}
Introducing the effective number of relativistic degrees of freedom, $\Neff$,
this expression can be re-written as follows ~\cite{Maggiore:1999vm,Mangano:2005cc}:
\begin{equation} \label{eq:Neff}
%  \Omega_{rad}=\left[1+\frac{7}{8}\Neff\left(\frac{T_{\nu}}{T_{\gamma}}\right)^4 \right] \Omega_{\gamma} \ .
  \Omega_{rad}=\left[1+\frac{7}{8}\Neff(\frac{4}{11})^{4/3} \right] \Omega_{\gamma} \ .
\end{equation}
%where the ratio of the temperatures of the neutrinos and the relic photons
%is given by $T_{\nu}/T_{\gamma}=(4/11)^{1/3}$ just after the neutrinos decoupling,
%assuming they are completely decoupled.
%where the factor of $7/8$ is due to the difference between the Fermi and
%Bose integrals  The ratio between the temperatures of relic photons
%and neutrinos $T_{\nu}/T_{\gamma}=(4/11)^{1/3}$ when assuming that
%neutrinos are not affected by the entropy transfered from
%electron-positron annihilations to photons.
Under the assumption that only photons and standard light neutrinos 
%sensitive
%to weak interactions 
contribute
to the radiation energy density, $\Neff$ is equal to the effective number
of neutrinos and is constrained by the measurement of the decay width of the
Z boson~\cite{Beringer:1900zz}. Taking into account the residual heating of the
neutrino fluid due to electron-positron annihilation,
its predicted value is $\Neff\simeq 3.046$~\cite{Mangano:2005cc}. Any deviation from this value can
be attributed to extra relativistic radiation, including massless
sterile neutrino species,
axions~\cite{Melchiorri:2007cd,Hannestad:2010yi}, decay of
non-relativistic matter~\cite{Fischler:2010xz}, GWs~\cite{Smith:2006nka}, extra
dimensions~\cite{Binetruy:1999hy,Shiromizu:1999wj,Flambaum:2005it},
early dark energy~\cite{Calabrese:2011hg}, asymmetric dark matter~\cite{Blennow:2012de}, or leptonic
asymmetry~\cite{Caramete:2013bua}. Assuming that a
stochastic CGW background contributes to this extra relativistic
radiation energy density, one can set an upper limit on the CGW background energy density,
$\Omegagw$, using Eq.~\ref{eq:Neff}:
\begin{equation} \label{eq:rhox}
  \Omegagw \leq \frac{7}{8}(\frac{4}{11})^{4/3}(\Neff-3.046)\Omega_{\gamma} \ .
\end{equation}
Using $T=2.7255\pm0.0006$~K for the photon temperature
today~\cite{Fixsen:2009ug}, one can deduce,
from~\cite{Beringer:1900zz}, the numerical value:
\begin{equation} \label{eq:h2}
  \Omega_{\gamma}   = 2.473 \ 10^{-5} /h_0^2 \ .
\end{equation}
Eq.~\ref{eq:rhox} then writes:
\begin{equation} \label{eq:NS}
  \Omegagwhdeux \leq  5.6 \ 10^{-6} (\Neff-3.046)\ .
\end{equation}
In the next section, we will measure $\Neff$ using the most recent
cosmological data and Eq.~\ref{eq:NS} will be used to constrain the
CGW background energy density.

\section{New constraints from Planck} \label{sec:planck}

\subsection{Data, models and statistical strategy} \label{sec:planck:data}
CMB and BAO data have been used in the following analysis.
The CMB data are composed of the temperature anisotropy
likelihood~\cite{Ade:2013kta} (hereafter
called Planck), and the CMB
lensing from Planck~\cite{Ade:2013tyw} (Lensing), WMAP
low-$\ell$ polarization (WP), SPT and
ACT~\cite{Reichardt:2011fv,Das:2013zf} high-$\ell$ spectra (HighL) data.
We combine them with BOSS BAO DR9~\cite{Beutler:2011hx,Padmanabhan:2012hf,Anderson:2012sa} data (BAO). In
addition, the robustness of the results has been checked 
against BOSS BAO DR11~\cite{Anderson:2013oza} data.

Profile likelihoods $\cal{L}$ of the $\Neff$ parameter have been
built; they are obtained by fixing $\Neff$ to predefined 
values, and maximizing the likelihood function over all the other
dimensions of the parameters space. 
The combined likelihood includes a total of 37 cosmological and
nuisance parameters as explained in~\cite{Planck:2013nga}.  
The maximisation is done using the Minuit software~\cite{James:310399},
interfaced to the
\textit{CLASS} Boltzmann solver~\cite{Blas:2011rf}, which, from a set of
input cosmological parameters, computes the corresponding CMB temperature
and polarization anisotropy power spectra. The tuning of the \textit{CLASS} precision
parameters has been done according to~\cite{Planck:2013nga}.  
Finally, a limit on \Omegagwhdeux\ using Eq.~(\ref{eq:NS}) is 
computed using the Feldman-Cousins prescription~\cite{Feldman:1997qc}
to take into account the condition $\Omegagwhdeux\ge0$.

Since the $\Omegagwhdeux$ limit is deduced from the $\Neff$
measurement, several neutrino models have been investigated for, in
order to study the sensitivity of our results to this unknown. Three
cases have been considered as (extended-)$\lambdaCDM$ scenarios: the
sum of the neutrino masses ($\sum{m_\nu}$) has either been set to 0, or assumed to be
0.06~eV as in~\cite{Ade:2013zuv,Planck:2013nga} which is
consistent with~\cite{Tortola:2012te}, or assumed to be free to vary and fitted for
for in the statistical procedure.

\subsection{Results}\label{sec:planck:results}

The measured 95$\%$ CL upper limits are summarized in Table~\ref{tab:limits}.
We have checked that the underlying $\Neff$ measurement is fully
compatible with~\cite{Ade:2013zuv}. For instance, for the Planck+WP+HighL+BAO
dataset and assuming $\sum{m_\nu}=0.06$~eV, the profile likelihood method
gives $\Neff=3.29\pm0.54$ at a 95$\%$ CL, very close to the value $3.30^{+0.54}_{-0.51}$
of ~\cite{Ade:2013zuv}. This shows that the results from both analyses using
different Bolzman solvers and different statistical methods are equivalent.

%For instance, for
%the case of $\sum m_\nu=0.06$~eV the minimum of the $-2\log\cal{L}$ is
%higher but the width of the distribution is smaller.
\begin{figure}%[tc]
  \includegraphics[width=0.5\textwidth]{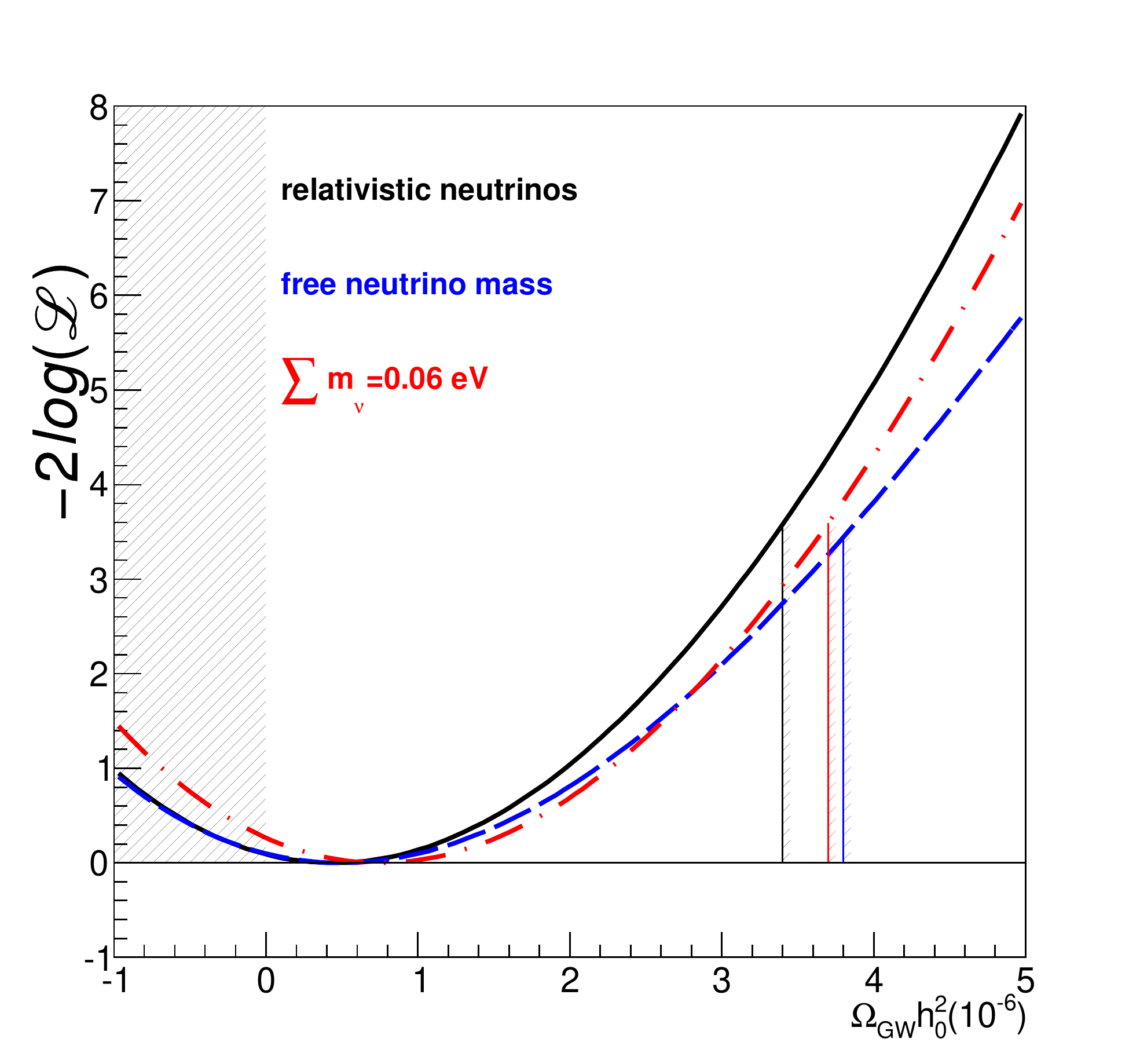}
  \centering
  \caption{Profile likelihoods of \Omegagwhdeux\ with the full Planck+WP+HighL+BAO+Lensing dataset (see Section~\ref{sec:planck:data} for details) in three neutrino models:
    when the neutrinos are considered to be massless (in 
    black full line), when the sum of the neutrino masses is assumed to
    be $\sum{m_\nu}=0.06$~eV (in red dashed and dotted line), and when the neutrino 
    mass is let free to vary in the fit (in blue dahed line). The grey area correspond
   to the unphysical $\Omegagwhdeux\le 0$ region. The limits
    are then derived using the Feldman-Cousins
    prescription and shown as vertical lines. }
  \label{fig:gw}
\end{figure}

Fig.~\ref{fig:gw}
shows the $-2\log\cal{L}$ functions for the full dataset, for the three
hypotheses of the neutrino masses discussed above.
The corresponding limits are reported in
Table~\ref{tab:limits}. The results are very robust to the considered neutrino models,
since fitting the $\sum{m_\nu}$ parameter does not significantly change the  upper limits.
In addition, the use of the DR11 BAO data in place of the DR9 does not 
improve the limit. On the other hand, CMB lensing data are very useful for improving the
constraints as the upper limit gets tighter by almost 20\%.

In the following, we consider the limit
obtained with the full dataset and with $\sum{m_\nu}$ being a free
parameter,
\begin{equation}
 \Omegagwhdeux < 3.8\times 10^{-6} \  \ .
\end{equation}

\begin{table}%[b!]
  \centering
  \begin{tabular}{|l|c|c|}
    \hline
    data                         & model             & 95$\%$CL Limits  \\
    \hline
    Planck+WP+HighL+BAO          & \multirow{2}{*}{$\sum{m_\nu}=0$}    & $<\ 4.1 \times 10^{-6}$ \\
    Planck+WP+HighL+BAO +Lensing         & & $<\ 3.4 \times 10^{-6}$ \\
   \hline
    Planck+WP+HighL+BAO          & \multirow{3}{*}{$\sum{m_\nu}=0.06$~eV}  & $<\ 4.3 \times 10^{-6}$ \\
    Planck+WP+HighL+BAO+Lensing  & & $<\ 3.7 \times 10^{-6}$ \\
    Planck+WP+HighL +BAODR11+Lensing & & $<\ 3.7 \times 10^{-6}$ \\
    \hline
%    Planck+WP+HighL              & $\sum(m_\nu)$ free & $<\ 5.4 \times 10^{-6}$ \\
    Planck+WP+HighL +BAO         & \multirow{2}{*}{$\sum{m_\nu}$ free} & $<\ 4.1 \times 10^{-6}$ \\
    \textbf{Planck+WP+HighL +BAO+Lensing} & & $\mathbf{<\ 3.8 \times 10^{-6}}$ \\
    % \hline
 %   Planck+WP+HighL +BAO+Lensing & $r=0.2$ and$\sum(m_\nu)=0.06$ & $<\ 5.\times 10^{-6}$ \\
    \hline
  \end{tabular}
  \caption{95$\%$ CL limits on $\Omegagwhdeux$ for different datasets
    and different neutrino mass hypothesis ($\sum{m_\nu}$ free means
    that the sum of the neutrino masses is a free parameter in the
    profile likelihood determination). Those limits are valid for
    frequencies above $10^{-15}$~Hz. The bold value is the one used
    in the following sections.}
  \label{tab:limits}
\end{table}
 
These upper limits can be compared to the ones previously derived in~\cite{Smith:2006nka}
and~\cite{Sendra:2012wh}. In the analysis of~\cite{Smith:2006nka}, the
quoted limit for adiabatic initial condition is $4\times 10^{-5}$ at 95$\%$~CL. In their analysis,
they assumed that the contribution of the standard neutrino only-component of $\Neff$ is 
$3.04$ and the neutrino masses were free to vary. This result is directly comparable
to our limit when fitting for $\sum{m_\nu}$, and  is an
order of magnitude larger. A more recent
result~\cite{Sendra:2012wh} gave $\Omegagwhdeux < 8.7 \times
10^{-6}$  at a 95$\%$~CL assuming massless neutrino species as is done
for the results presented in the first lines of Table~\ref{tab:limits}, leading to a result more than a
factor two above our final results. They also quote a limit of $10^{-6}$ for homogeneous
initial conditions which are not investigated here.
Apart from the fact that the limit
obtained in this paper
is the lowest as of today for adiabatic initial conditions,
we have also shown that it is robust to the assumptions
made on the neutrino mass.

\section{Constraints on cosmic strings models}\label{sec:cs}

The limit on the primordial stochastic GW background can be used to
constrain cosmic string models, for if cosmic strings exist, they would be
a powerful source of GWs. Bursts of GWs are emitted by string
features called cusps and kinks that propagate along string loops,
a result of cosmic string interactions. The GWs emitted by cosmic 
strings are described by the dimensionless string
tension parameter $G\mu$, where $G$ is the Newton constant and $\mu$
the mass per unit length of the string. Recent results from
Planck~\cite{Ade:2013xla} showed that, in order to be compatible with
the measured CMB anisotropies, the string tension had to be lower than
$1.5\times 10^{-7}$ and $3.2\times 10^{-7}$ for Nambu-Goto and
Abelian-Higgs strings respectively.  When two string segments meet,
they have a probability to exchange ends; strings are said to
intercommute or to reconnect. A network of cosmic strings is then also
characterized by a reconnection probability, $p$. For topological
strings, $p$ is essentially 1 while it can be much smaller for cosmic
strings formed in the context of string theory
(superstrings)~\cite{Jackson:2004zg}. When a single string reconnects
with itself, a closed loop breaks off, oscillates and radiates
gravitationally through the formation of cusps and kinks. This
mechanism was first described in~\cite{Damour:2000wa} and has
been extended more recently in~\cite{Olmez:2010bi}. For clarity, the
main analysis steps of~\cite{Olmez:2010bi}, leading to constraints on
cosmic string models, are summarized below.

When considering an incoherent superposition of GWs emitted by cusps
or kinks, the GW spectrum, defined in Section~\ref{sec:introduction}, is
given by
\begin{equation}\label{eq:Omega_cs}
  \Omegagw(f)=
  \frac{4 \pi^2}{3 H_0^2}f^3
  \int{dz \int{dl h^2(f,z,l)\frac{d^2R(z,l)}{dzdl}}}.
\end{equation}
The GW strain amplitude produced by a cusp or a kink at a redshift $z$ is
represented by $h$. The GW burst rate per loop length per redshift is
$\frac{d^2R(z,l)}{dzdl}$. The loop size is typically taken as a
fraction of the horizon size at cosmic time $t$: $l\sim\alpha
t$. Early simulations of cosmic string evolution, such as~\cite{Bennett:1987vf}, suggested that
the loop size was determined by the gravitational back-reaction, in
which case loops are short-lived and $\alpha\leq\Gamma G\mu$, where
$\Gamma\sim50$ is related to the fraction of power going into
GWs. Recent simulations~\cite{Ringeval:2005kr,Blanco-Pillado:2013qja}
showed that the gravitational back-reaction scale is irrelevant and the
formation of large loops is favored ($\alpha\sim 0.05$). The
small and large loop scenarios strongly impact the loop density
that is used to compute the expected GW burst rate. In the
following we examine both cases and derive the corresponding upper
limits using the constraint obtained in Section~\ref{sec:planck}.

%%%%%%%%%%%%%%%%%%%%%%%%%%%%%%%%%%%%%%%%%%%%%%%%%%%%%%%%%%%%%%%%%%%%%%%%
%%%%%%%%%%%%%%%%%%%%%%%%       SMALL LOOP       %%%%%%%%%%%%%%%%%%%%%%%%
%%%%%%%%%%%%%%%%%%%%%%%%%%%%%%%%%%%%%%%%%%%%%%%%%%%%%%%%%%%%%%%%%%%%%%%%
%\subsection{}\label{sec:cs:small}
If loop sizes at the time of formation were driven by the gravitational
back-reaction, the rate of GW bursts would be computed using a loop density

\begin{equation}\label{eq:density_small}
  n(l,t)=c(z)(p\Gamma G\mu)^{-1}t^{-3}\delta(l-\alpha t),
\end{equation}
where $c(z)=1+9z/(z+z_{eq})$~\cite{Damour:2000wa,Olmez:2010bi} and
$z_{eq}=3366$~\cite{Ade:2013zuv} is the redshift at matter-radiation equality. This
expression simplifies the integral over $l$ in Eq.~\ref{eq:Omega_cs} to
\begin{equation}\label{eq:spec_small_cusp}
  \Omegagw^{cusp}(f)=
  \frac{2 G\mu \pi^2 H_0^{1/3}}{3 p \alpha^{1/3} \Gamma f^{1/3}}
  \times \int{ dz \frac{c(z) \varphi_V(z) \Theta\left[1-\left(f(1+z)H_0^{-1} \alpha \varphi_t(z)\right)^{-1/3}\right]}{(1+z)^{7/3}\varphi_r^2(z)\varphi_t^{10/3}(z)}},
\end{equation}
for cusps and
\begin{equation}\label{eq:spec_small_kink}
  \Omegagw^{kink}(f)=
  \frac{4 G\mu \pi^2 H_0^{2/3}}{3 p \alpha^{2/3} \Gamma f^{2/3}}
  \times \int{ dz \frac{c(z) \varphi_V(z) \Theta\left[1-\left(f(1+z)H_0^{-1} \alpha \varphi_t(z)\right)^{-1/3}\right]}{(1+z)^{8/3}\varphi_r^2(z)\varphi_t^{11/3}(z)}},
\end{equation}
for kinks, where we used the gravitational waveform $h$ and the rate
$\frac{d^2R(z,l)}{dzdl}$ modeled in~\cite{Olmez:2010bi}. The
expressions differ because kinks and cusps have
different waveforms and different GW beaming
angles. Three dimensionless functions, $\varphi_t(z)$, $\varphi_r(z)$
and $\varphi_V(z)$, are used to reflect the redshift dependence
of cosmic time, proper distance and volume,
respectively. These cosmological functions are numerically
computed using the expressions defined in~\cite{Olmez:2010bi} and the
cosmological parameters from~\cite{Ade:2013zuv}.

\begin{figure}
  \center
  \includegraphics[width=0.7\textwidth]{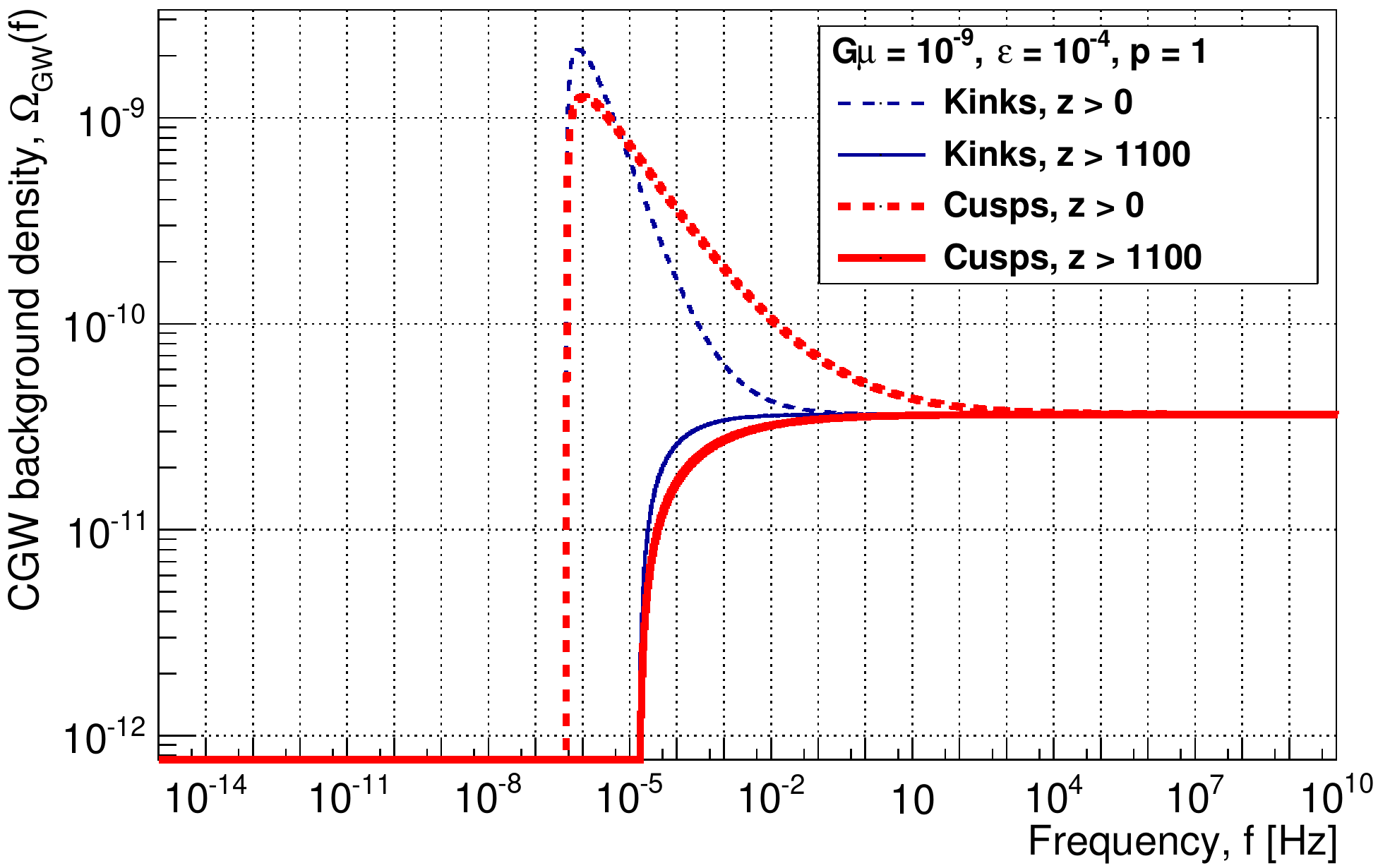}
  \caption{Kink (thin blue lines) and cusp (thick red lines) predicted
    spectra for small cosmic string loops ($\alpha \ll \Gamma
    G\mu$) using $G\mu=10^{-7}$, $\varepsilon=10^{-4}$ and
    $p=1$. These spectra are derived numerically from
    Eq.~\ref{eq:spec_small_cusp} and Eq.~\ref{eq:spec_small_kink}
    using $z>1100$ for GWs produced prior to the photon decoupling and
    $z>0$ for GWs produced up to today.}
  \label{fig:spectrum}
\end{figure}

\begin{figure}
  \center
  \includegraphics[width=0.49\textwidth]{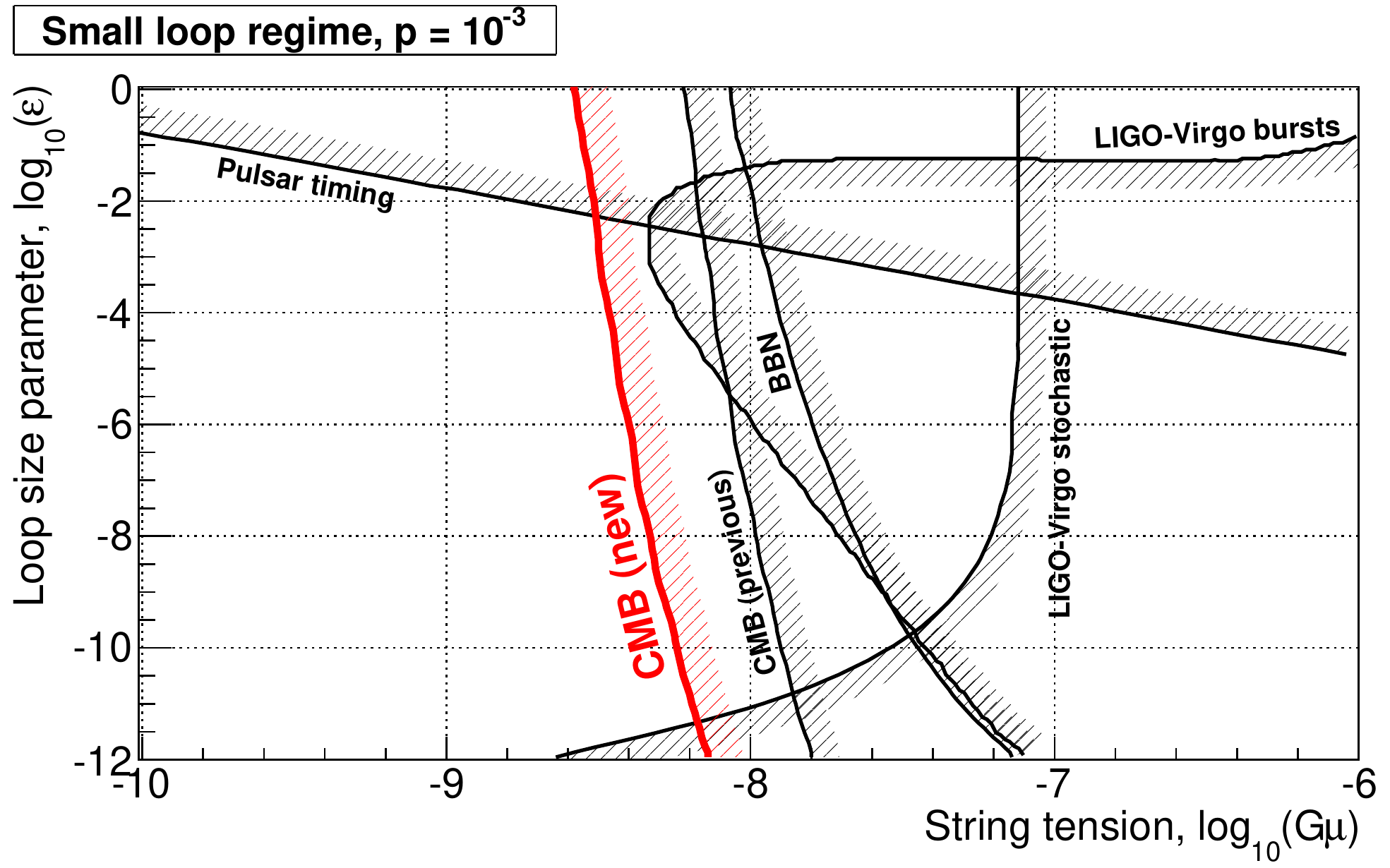}
  \includegraphics[width=0.49\textwidth]{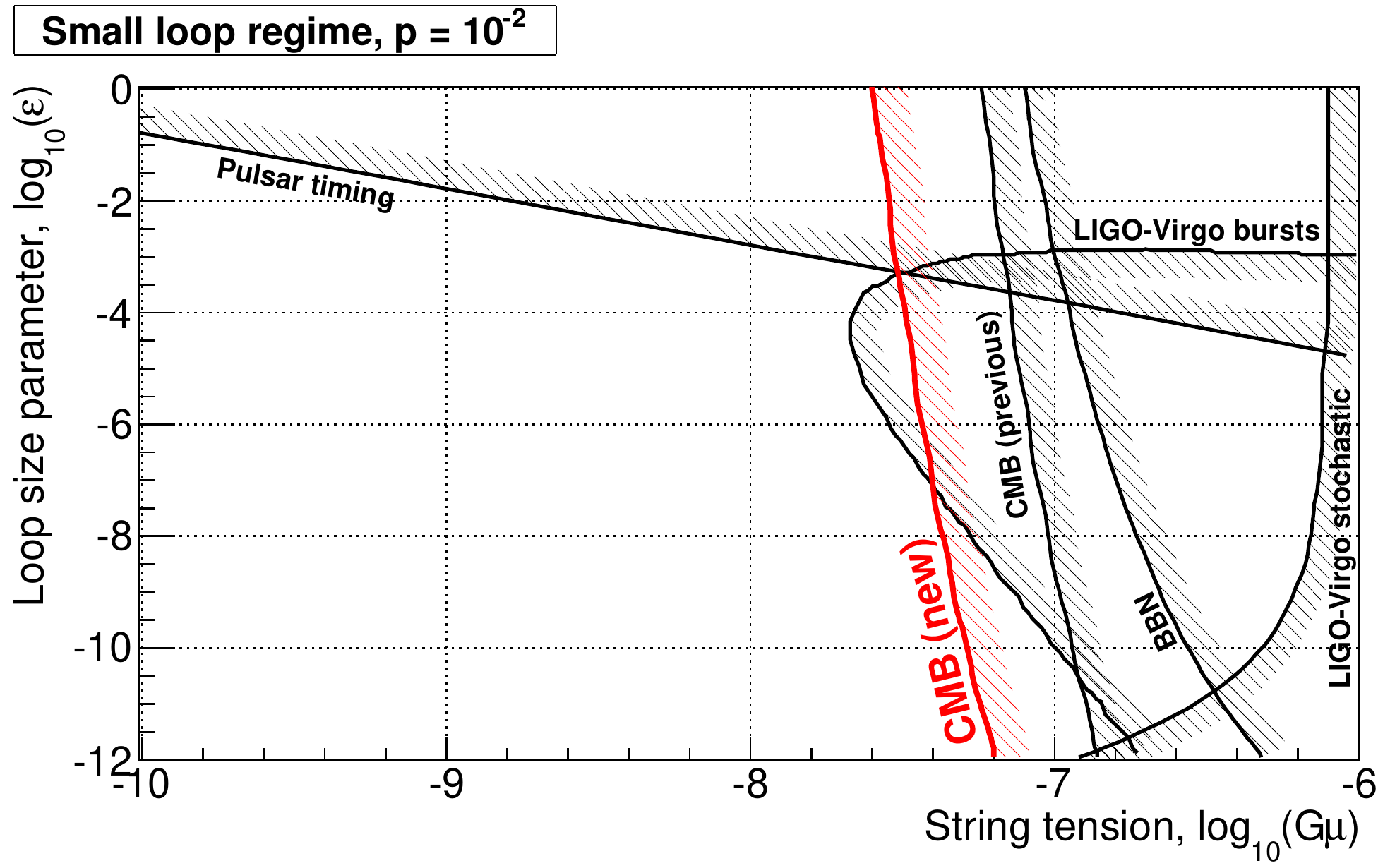} \\
  \includegraphics[width=0.49\textwidth]{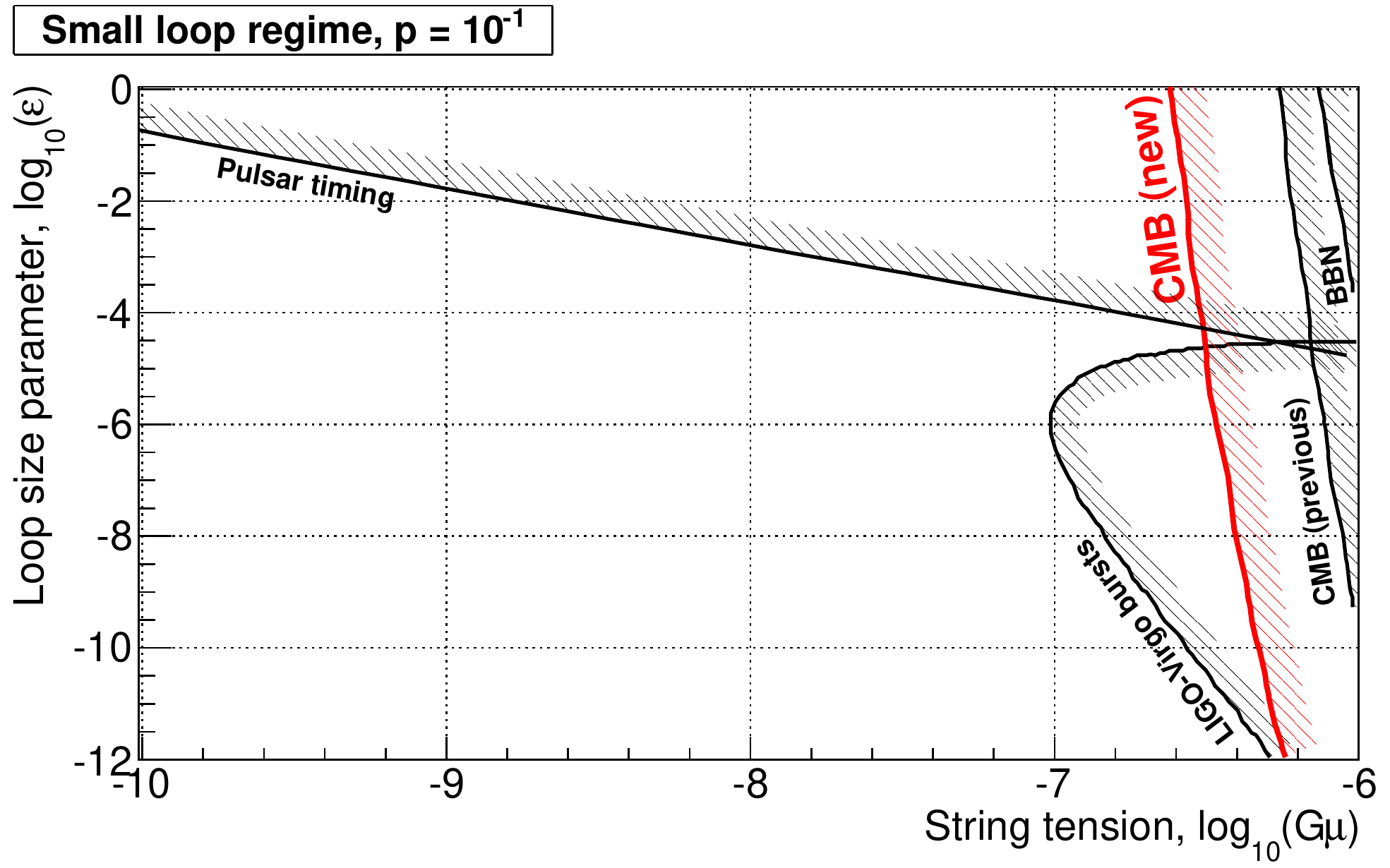}
  \includegraphics[width=0.49\textwidth]{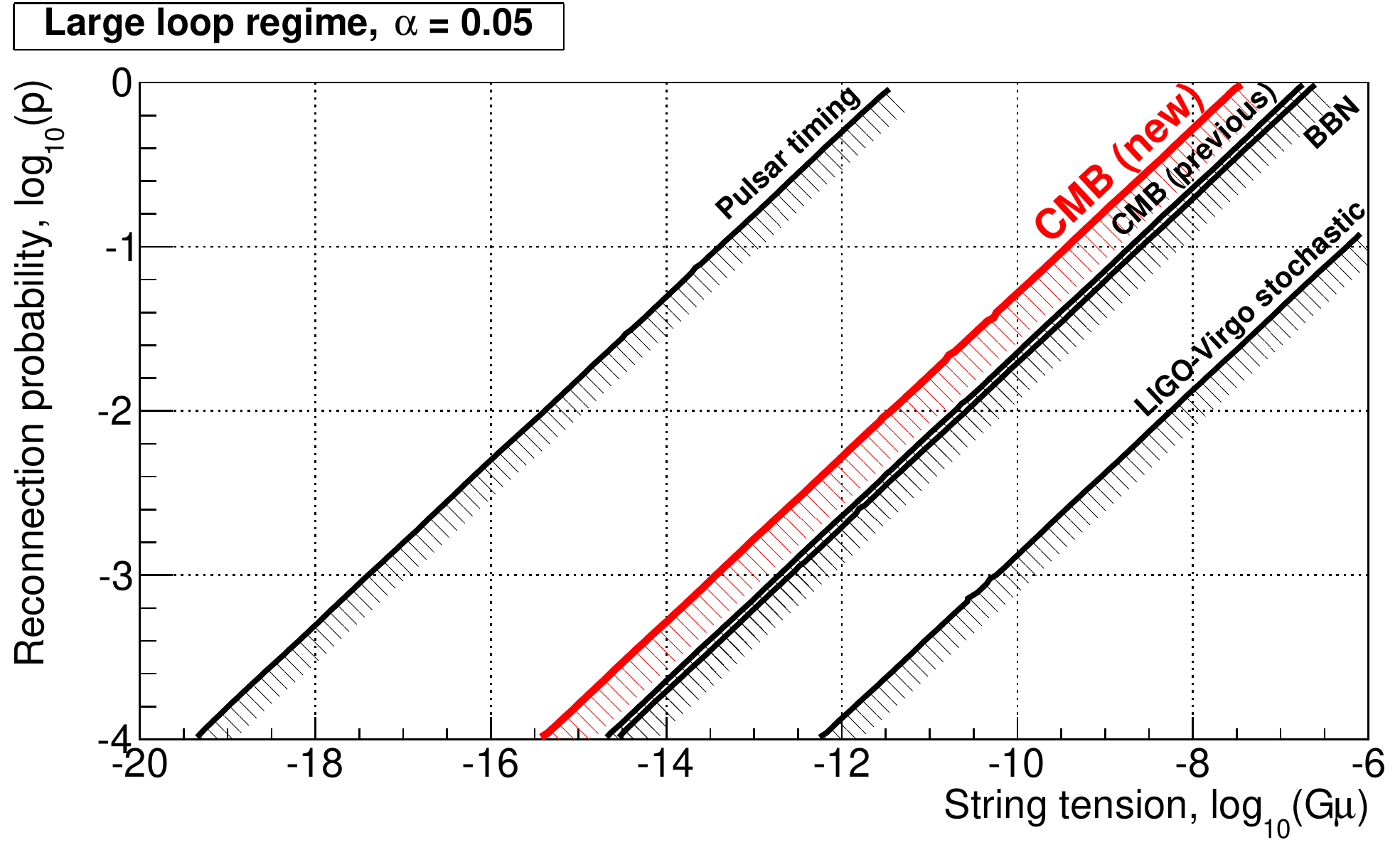}
  \caption{Constraints on cosmic string parameters
    $(G\mu,\varepsilon)$ assuming loops are small and fixing
    $p=10^{-3}$ (top-left), $p=10^{-2}$ (top-right), and $p=10^{-3}$
    (bottom-left). The lower-right plot shows exclusion contours in
    the $(G\mu,p)$ plane when loops are large and for $\alpha=0.05$~\cite{Blanco-Pillado:2013qja}. Our
    new constraints, derived from $\Omegagwhdeux < 3.8\times 10^{-6}$, are
    compared to existing limits obtained from previous CMB
    ($\Omegagwhdeux < 8.7 \times 10^{-6}$~\cite{Sendra:2012wh}), BBN
    ($\Omegagwhdeux < 8.1 \times 10^{-6}$~\cite{Cyburt:2004yc}), LIGO-Virgo
    stochastic ($\Omegagw(f\simeq 100\ \text{Hz})\times
    (\hzero/0.68)^2<5.6\times 10^{-6}$~\cite{Aasi:2014zwg}),
    LIGO-Virgo bursts~\cite{Aasi:2013vna} and pulsar timing
    ($\Omegagw(f=2.8\ \text{nHz})\times (\hzero/0.73)^2 < 1.3 \times
    10^{-9}$~\cite{Shannon:2013wma}) data analyses.}
  \label{fig:ul}
\end{figure}

For this study, we compute the GW spectrum for kinks and
cusps using a set of cosmic string parameters $(G\mu, \varepsilon,
p)$, where $\varepsilon\equiv\alpha/(\Gamma
G\mu)$, with $\varepsilon < 1$, is another parameterization of the
loop size. Examples of such spectra are
presented in Fig.~\ref{fig:spectrum}. The cosmic string parameter
space is scanned and, for each parameter set, we compute the
spectrum integral
$\int{d(\ln{f})(\Omegagw^{cusp}(f)+\Omegagw^{kink}(f))}$. The
integration is performed from $10^{-15}$~Hz up to $10^{10}$~Hz, to
encompass GWs produced after the phase transition which produced the
string network and before the time of the photon decoupling, and using
$h_0=0.6780\pm 0.0077$~\cite{Ade:2013zuv}. Cosmic string parameters
are excluded by requiring this integral to be smaller than the upper
limit on $\Omegagw$ obtained in Section~\ref{sec:planck}. The upper
plots and the lower-right plot of Fig.~\ref{fig:ul} present the
resulting exclusion contours at 95\%~CL (in red and labeled ``CMB new'')
when $p$ is fixed at $10^{-3}$, $10^{-2}$ and $10^{-1}$.

Fig.~\ref{fig:ul} also displays existing constraints which have been
derived using the model of~\cite{Olmez:2010bi} and rescaled with a
uniform set of cosmological parameters~\cite{Ade:2013zuv}. Our new
constraint on the string tension is
tighter by more than a factor 2 (resp. a factor 4) as compared to
previous results using CMB data~\cite{Sendra:2012wh} (resp. BBN
data~\cite{Cyburt:2004yc}). When compared to LIGO-Virgo results from
searches for a stochastic GW background~\cite{Aasi:2014zwg}, our new
results are an improvement by about an order of magnitude. We also
consider the recent search for individual GW bursts from
cusps~\cite{Aasi:2013vna,Siemens:2006vk} compared to which our constraints does not
scale the same way with $p$ in the $(G\mu,\varepsilon)$ plane: our new
bound is only competitive for $p\lesssim 10^{-2}$. Finally, we have
computed the constraints from pulsar timing data using the limit on
$\Omegagw(f)$ from~\cite{Shannon:2013wma}; it complements our result
for $\varepsilon\gtrsim 10^{-3}$. It is important to note that for
both the pulsar timing and the LIGO-Virgo results, the constraints are
not exactly comparable as these results also include GWs produced
after the photon decoupling\footnote{The $z$ integrations of
  Eq.~\ref{eq:spec_small_cusp} and Eq.\ref{eq:spec_small_kink} are
  performed using $z>1100$ for CMB, $z>5.5\times 10^{9}$ for BBN,
  and $z>0$ for LIGO-Virgo stochastic and pulsar timing results.}.

We also examine the case where loops are large and their size is set
by the large-scale dynamics of the string network, as suggested by
recent simulations~\cite{Ringeval:2005kr,Blanco-Pillado:2013qja}. The
loop size is no longer parameterized by $\varepsilon$; $\alpha$ is
used instead. We use the analytical approximation derived
in~\cite{Olmez:2010bi} where the CGW spectrum is flat for the
radiation era ($z>z_{eq}$). This result is obtained using a loop
density, $n(l,t)$, specifically modeled for the radiation
era~\cite{Siemens:2006vk} and integrated over in
Eq.~\ref{eq:Omega_cs}. For both cusps and kinks the CGW spectrum is
approximated by 
\begin{equation}\label{eq:spec_large} 
  \Omegagw(f) = \frac
          {192\pi^3 \sqrt{\alpha\Gamma G\mu}}
          {3z_{eq}\Gamma^2p} \ .
\end{equation}
The lower-right plot of Fig.~\ref{fig:ul} shows how the $G\mu-p$
plane is constrained when fixing $\alpha$ at 0.05~\cite{Blanco-Pillado:2013qja}. As expected when using
a $f$-independent spectrum, the best constraint is provided by the
tightest limit on $\Omegagw(f)$, independently of the frequency
band. In this context, the pulsar timing constraint is the most
competitive one.

\section{Conclusions} \label{sec:conclusions}

Assuming that the deviation of \Neff\ from its expected value is due to the 
CGW background, we have analyzed the recent CMB 
(from Planck, WMAP low-$\ell$ polarization, SPT and ACT~\cite{Ade:2013kta,Ade:2013tyw,Reichardt:2011fv,Das:2013zf}) and 
BOSS BAO DR9 data~\cite{Beutler:2011hx,Padmanabhan:2012hf,Anderson:2012sa} assuming adiabatic initial conditions.
Using a profile  likelihood method we have derived an upper limit on the CGW energy
density for frequencies greater than $10^{-15}$~Hz : $\Omegagwhdeux < 3.8
\times 10^{-6}$ at 95\% confidence level. With $\hbox{h}_0=0.678$~\cite{Ade:2013zuv} this implies  $\Omegagw < 8.3
\times 10^{-6}$.
This result is a factor 2.3
better than previous ones~\cite{Smith:2006nka,Sendra:2012wh}, and
is robust to
the neutrino mass models. 
We have shown that adding
the lensing
CMB data to the Planck temperature measurements significantly improves the constraints. On the contrary, the use of 
DR11 BAO data~\cite{Anderson:2013oza} does not add enough information to significantly change
the upper limit.

Under the hypothesis that the primordial GWs can be attributed to a
network of cosmic strings, we have computed exclusion regions in the
string parameter space for any possible size of loops. The best limit
is obtained in the scenario where the size of loops is dictated by the GW
back-reaction. In particular, in the context of string theory, if the
cosmic string reconnection probability is $10^{-3}$, we exclude models
with $G\mu \gtrsim 4\times 10^{-9}$. For the large loop case, the best
limits are still provided by the pulsar timing experiments~\cite{Shannon:2013wma}.

Our upper bound on the adiabatic energy density and the BBN limit
($\Omegagwhdeux < 8.1 \times 10^{-6} $~\cite{Cyburt:2004yc,Abbott:2009ws}) covers a very large
range of frequencies, $10^{-15}$ and $10^{-10}$ Hz and above respectively, while the direct
search limits set with LIGO-Virgo data ($\Omegagw(f\simeq 100\ \text{Hz}) \times (\hzero/0.68)^2 < 5.6\times
10^{-6}$~\cite{Aasi:2014zwg}) and  pulsar timing experiments ($\Omegagw(f=2.8\ \text{nHz}) \times (\hzero/0.73)^2
< 1.3 \times 10^{-9}$~\cite{Shannon:2013wma}) are obtained over a much smaller frequency band. 
The sensitivity of the next generation of interferometric GW
detectors will improve by a factor 10, especially at low frequencies (down
to 10~Hz). With this sensitivity, the GW background energy density constraint will be of the order
of $\Omegagw(f\simeq 10-150\ \text{Hz})\lesssim 10^{-9}$ (assuming a year of
data at LIGO-Virgo design sensitivity)~\cite{Aasi:2014zwg}.

%Recently the Bicep 2 team~\cite{Ade:2014xna} disfavored the null value of the
%tensor to scalar ratio $r$ at 7$\sigma$ through the detection
%of CMB B polarization modes. These tensor perturbations 
%suggesting the existence of a stochastic gravitational wave 
%background need to be confirmed by future experiments
Future Planck  measurements using polarization may slightly reduce the
error on \Neff. 
In the coming years, ground based or balloon borne CMB observation projects should also be able to improve on this
measurement~\cite{Ade:2014xna,Ade:2014afa,2014arXiv1407.5928A,2012SPIE.8452E..1EA,2010SPIE.7741E..1SN,2010SPIE.7741E..1CR,2010SPIE.7741E..1NF}. In the long term,
Euclid~\cite{Amendola:2012ys} or LSST~\cite{Kitching:2008dp} will reduce the error on \Neff\ down to
0.1 at 1$\sigma$, a factor 2 smaller than
the current error.
% What is meant by the actual error.
Still, the prediction of these sensitivities in terms
of achievable upper limits on the CGW background energy density
depends highly on the $\Neff$ value itself within the 
hypotheses assumed in this paper.
More constraints should come in ten years from now, from high precision ground-based
instruments measuring the CMB polarization anisotropies. For example the forecasted stage-IV
CMB polarization experiment CMB-S4 should allow one to achieve
an error on \Neff\ of 0.02 at 68$\%$~CL~\cite{Abazajian:2013oma}, i.e. more than
one order of magnitude smaller that was has been attained so
far. Moreover, the improvement on the sensitivity of the 
tensor-to-scalar ratio (of the order of $\simeq 0.001$ with a similar systematic
error~\cite{Abazajian:2013vfg}), would provide a confirmation of the existence of the
inflationary cosmological GW background, if any.

%Others cosmological probes like supernovae might also help
%reducing the global error on the \Neff measurements with different
%type of systematic errors.

\section*{Acknowledgements}
\noindent NC's work is supported by NSF grant PHY-1204371. We thank Vuk Mandic for useful conversations.

\bibliographystyle{iopart-num}
\section*{References}
\bibliography{references}

\end{document}